\documentclass[sigconf, nonacm=true]{acmart}

\usepackage{algorithm}
\usepackage{algorithmic}
\usepackage{graphicx}
\usepackage{textcomp}
\usepackage{amsmath}
\usepackage{array}
\usepackage{booktabs}
\usepackage{xspace}

\newcommand{\cohmeleon}[0]{{\sc cohmeleon}\xspace}

\AtBeginDocument{%
  \providecommand\BibTeX{{%
    \normalfont B\kern-0.5em{\scshape i\kern-0.25em b}\kern-0.8em\TeX}}}

\copyrightyear{2021}
\acmYear{2021}
\setcopyright{none}\acmConference[MICRO '21]{MICRO-54: 54th Annual
IEEE/ACM International Symposium on Microarchitecture}{October 18--22,
2021}{Virtual Event, Greece}
\acmBooktitle{MICRO-54: 54th Annual IEEE/ACM International Symposium on
Microarchitecture (MICRO '21), October 18--22, 2021, Virtual Event, Greece}
\acmPrice{15.00}
\acmDOI{10.1145/3466752.3480065}
\acmISBN{978-1-4503-8557-2/21/10}

\acmConference[MICRO'21]{The 54th Annual IEEE/ACM International Symposium on
Microarchitecture}{October 18-22, 2021}{Athens, Greece}

\newcommand{\rev}[1]{{#1}\xspace}
\newcommand{\cam}[1]{{#1}\xspace}

\makeatletter
\def\blfootnote{\xdef\@thefnmark{}\@footnotetext}
\makeatother

\title{
    Cohmeleon: Learning-Based Orchestration of \\
    Accelerator Coherence in Heterogeneous SoCs
}

\begin{document}
\title{
    Cohmeleon: Learning-Based Orchestration of \\
    Accelerator Coherence in Heterogeneous SoCs
}

\author{Joseph Zuckerman}
\email{jzuck@cs.columbia.edu}
\affiliation{%
  \institution{Department of Computer Science, Columbia University}
  \city{New York}
  \state{New York}
  \country{USA}
  \postcode{10027}
}

\author{Davide Giri}
\email{davide\_giri@cs.columbia.edu}
\affiliation{%
  \institution{Department of Computer Science, Columbia University}
  \city{New York}
  \state{New York}
  \country{USA}
  \postcode{10027}
}

\author{Jihye Kwon}
\email{jihyekwon@cs.columbia.edu}
\affiliation{%
  \institution{Department of Computer Science, Columbia University}
  \city{New York}
  \state{New York}
  \country{USA}
  \postcode{10027}
}

\author{Paolo Mantovani}
\email{paolo@cs.columbia.edu}
\affiliation{%
  \institution{Department of Computer Science, Columbia University}
  \city{New York}
  \state{New York}
  \country{USA}
  \postcode{10027}
}
\authornote{Paolo Mantovani is now with Google Research.}

\author{Luca P. Carloni}
\email{luca@cs.columbia.edu}
\affiliation{%
  \institution{Department of Computer Science, Columbia University}
  \city{New York}
  \state{New York}
  \country{USA}
  \postcode{10027}
}

\begin{abstract}
One of the most critical aspects of integrating \rev{loosely-coupled}
accelerators in heterogeneous SoC architectures is orchestrating their
interactions with the memory hierarchy, especially in terms of navigating the
various cache-coherence options\cam{: from accelerators accessing
off-chip memory directly, bypassing the cache hierarchy, to accelerators having
their own private cache}.
By running real-size applications on FPGA-based prototypes of
many-accelerator multi-core SoCs, we show that the best
cache-coherence mode for a given accelerator varies at runtime, depending on
the accelerator's characteristics, the workload size, and the overall SoC status.

Cohmeleon applies reinforcement learning to select the best coherence mode
for each accelerator dynamically at runtime, as opposed to statically at design
time.
It makes these selections adaptively, by continuously observing
the system and measuring its performance.
Cohmeleon is accelerator-agnostic, architecture-independent, and it requires
minimal hardware support. Cohmeleon is also transparent to application
programmers and has a negligible software overhead.
FPGA-based experiments show that our runtime approach offers, on average, a
38\% speedup with a 66\% reduction of off-chip memory accesses compared to
state-of-the-art design-time approaches. Moreover, it can match runtime
solutions that are manually tuned for the target architecture.
\end{abstract}

\ccsdesc[500]{Computer systems organization~System on a chip}
\ccsdesc[300]{Computer systems organization~Reconfigurable computing}
\ccsdesc[300]{Computer systems organization~Heterogeneous (hybrid) systems}
\ccsdesc[500]{Computing methodologies~Reinforcement learning}

\keywords{cache coherence, hardware accelerators, q-learning, system-on-chip}

\maketitle

\section{Introduction}
\label{sec:intro}

Modern computing systems of all kinds increasingly rely on heterogeneous
system-on-chip (SoC) architectures, which combine general-purpose processor
cores with many domain-specific hardware accelerators \cite{mobileye_ces18,
  xavier_hc18, xilinx_hc18, simba_micro19, cong_dac14, lyons_taco12,
  carloni_wcae19}.
In the case of smartphones, for example, major vendors
devote most of their SoC area to a growing number of
specialized hardware blocks~\cite{shao_mc15, alp_arxiv20}.
A multitude of accelerators have been designed
for many different application domains
\cite{nvdla, chen_jssc17, maeri_asplos18, scnn_isca17,
  eie_isca16, matraptor_micro20, sigma_hpca20, casper_fpga14, q100_asplos14,
  yao_micro20, graphpulse_micro20, seedex_micro20, edgebert_arxiv21}.

\rev{This work is focused on \emph{fixed-function loosely-coupled} accelerators (LCAs),
  a very common category of accelerators that includes, for instance, the NVIDIA
  Deep Learning Accelerator (NVDLA) -- part of the NVIDIA Tegra Xavier
  SoC~\cite{nvdla,xavier_hc18}.}
\blfootnote{To appear in the 54th IEEE/ACM
International Symposium on Microarchitecture (MICRO 2021)}
By following the loosely-coupled approach, these hardware accelerators are
designed independently from the processor cores, lie outside the core on the
system interconnect, \rev{are invoked through a device driver, execute
  coarse-grained tasks with no need for fine-grained synchronization}, and are
shared among multiple cores on an as-needed basis~\cite{chen_iccd13,
  cota_dac15}.
\rev{Fixed-function accelerators are not programmable (i.e. they do not execute
  instructions), but they can be highly configurable.}

\begin{figure}[t!]
\centering
\includegraphics[width=\columnwidth]{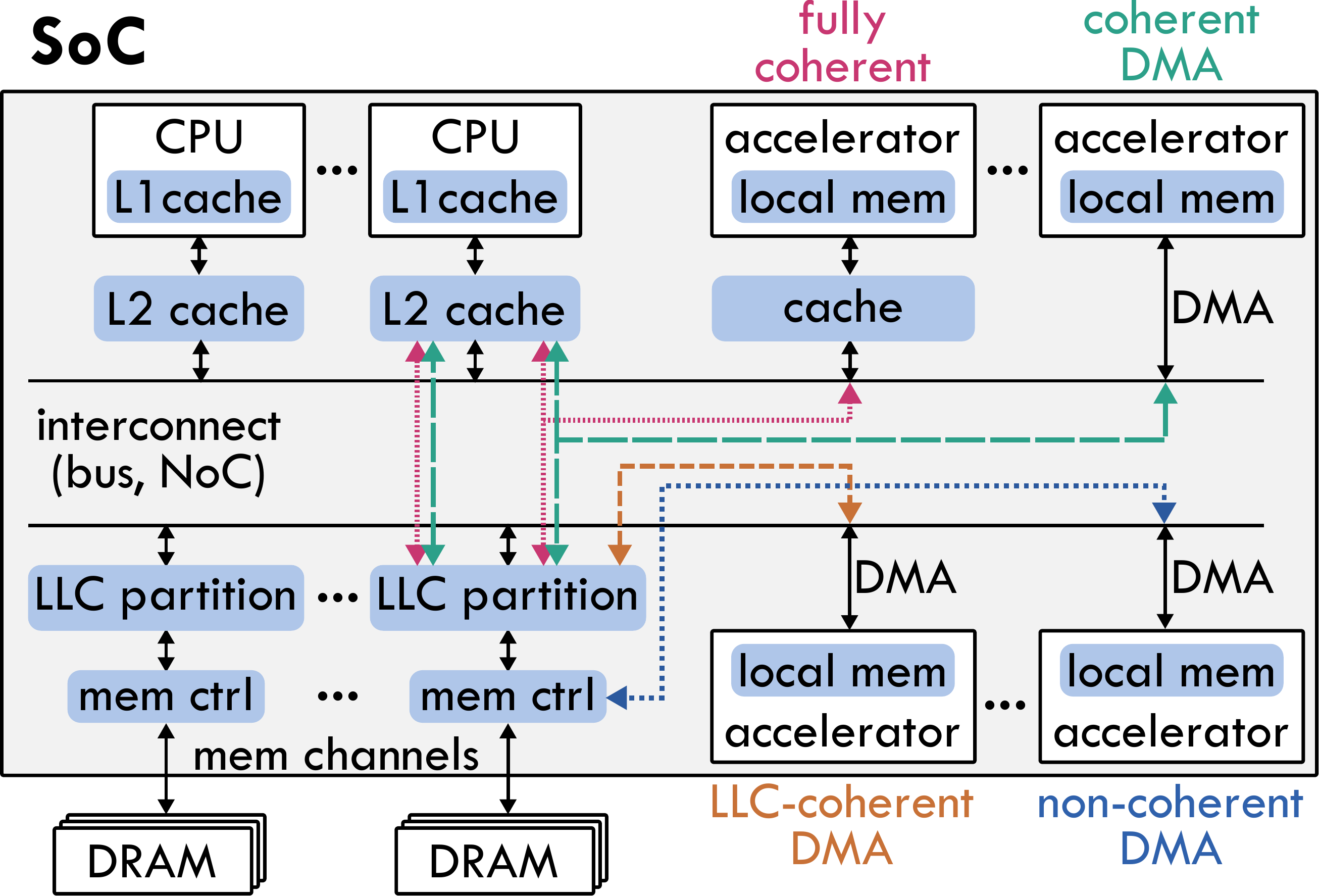}
\vspace{-0.25cm}
\caption{Memory hierarchy of a heterogeneous SoC with partitioned memory space
  and multiple DRAM controllers. The colored lines show the different ways
  accelerators can interact with the memory hierarchy.}
\label{fig:soc-memory}
\vspace{-0.5cm}
\end{figure}

For key computational tasks, hardware specialization delivers order-of-magnitude
gains in energy-efficient performance compared to software execution on
general-purpose processors~\cite{Horowitz2014}.
These gains are due to a combination of specialization and parallelism, but they
also require efficient memory subsystems~\cite{dally2020}.  Indeed, the
integration of \rev{LCAs} requires managing their interaction with the SoC
memory hierarchy, which is typically designed around the processor cores.
\figurename~\ref{fig:soc-memory} highlights the memory hierarchy of a generic
heterogeneous SoC, where a multi-level cache hierarchy supports the processor
cores' operation.
Complex SoCs may have a partitioned
last-level cache (LLC) as well as multiple DRAM controllers and channels
to increase the off-chip bandwidth and better distribute the
traffic~\cite{balkind_asplos16, blackparrot_ieeemicro20, mantovani_iccad20}.
LCAs interact directly with the SoC
memory hierarchy to load data into their private scratchpad and to store
results back to memory~\cite{lyons_taco12, cota_dac15, buffets_asplos19}.

\cam{Most LCAs process data in patterns that are very specific to the particular
  algorithm they implement. Since these patterns are predictable, designers
  exploit them to specialize the microarchitectures of the accelerator
  scratchpad and datapath~\cite{soldavini_arxiv21}. As a result, while some
  accelerators have irregular memory access patterns that resemble those of
  general-purpose cores, many others benefit from a different memory hierarchy,
  more tailored to their needs~\cite{shao_micro16, spandex_isca18,
    giri_ieeemicro18, buffets_asplos19}}.

\rev{The literature proposes many different modes for the interaction between
  LCAs and the memory hierarchy,} ranging from accelerators accessing off-chip
memory directly, bypassing the cache hierarchy, to accelerators having their own
private cache~\cite{fusion_isca15, giri_ieeemicro18, giri_nocs18, bhardwaj_islped20}.
\rev{Even though fixed-function LCAs do not require fine-grained synchronization
  like programmable accelerators (e.g. GPUs) or tightly-coupled accelerators
  normally do, they may still benefit from accessing on-chip cached data and
  from using hardware-based coherence provided by the cache hierarchy as an 
  alternative to software-based coherence mechanisms, such as cache flushes.}
By extending existing classifications, we identify four main
\emph{cache-coherence modes} for accelerators.

Then, we show that the best coherence mode varies at runtime, based on the
communication properties of the specific accelerator, its invocation parameters
(e.g. the size of the task it executes), and the overall system status (e.g. the
number of accelerators active in parallel). Thus, selecting an accelerator's
cache-coherence mode statically at design-time yields suboptimal performance
with respect to a runtime approach.
This is especially true for complex SoCs, where many concurrent applications
consisting of multiple threads that invoke multiple LCAs in
parallel to execute coarse-grained tasks. The on-chip traffic may vary
considerably depending on which accelerators are running and what
parameters they have been configured with.

For these reasons, we make the case that SoC architectures should support
multiple cache-coherence modes for accelerators as well as their runtime
selection. Accordingly, we propose \cohmeleon, a learning-based approach for
selecting the best coherence mode for \rev{fixed-function LCAs} dynamically at
runtime, as opposed to statically at design time.
\cohmeleon continuously observes the system and measures its performance by
means of a few hardware monitors and performance counters that are commonly
found in SoC architectures.  With this information, \cohmeleon trains a
reinforcement learning model to select the cache-coherence mode for each
\rev{fixed-function LCA} invocation. Our learning model has the flexibility to
target multiple optimization objectives simultaneously, like execution time and
off-chip memory accesses.
\cohmeleon is fully transparent to application programmers, and it adds a
negligible overhead to the accelerator invocations. Our approach does not
require any prior knowledge about the accelerators in the SoC and it is easily
applicable to different SoC architectures. \cohmeleon requires minimal hardware
support, provided that the target architecture has support for multiple
cache-coherence modes.

To evaluate \cohmeleon, we realized a set of complex FPGA-based prototypes of
SoC architectures. Each SoC has multiple LLC partitions and multiple DRAM
controllers for parallel access to main memory.  Using the FPGA as an
experimental infrastructure allows us to run real-size applications on top of
Linux SMP, without losing accuracy.
We developed a set of multi-threaded applications with broad coverage in terms
of accelerator parallelism and workload sizes.  The applications invoke a
variety of open-source accelerators as well as a traffic-generator that
reproduces the communication characteristics of a wide range of fixed-function
accelerators.
%

Our experiments show that \cohmeleon, on average, gives a speedup of 38\%, while
reducing the off-chip memory accesses by 66\% compared to state-of-the-art
design-time solutions. We also show that \cohmeleon, without any prior knowledge
on the target architecture, can match the performance of a runtime algorithm
manually tuned for it.

To build our SoC prototypes we leveraged ESP, an open-source platform for agile
design and prototyping of heterogeneous SoCs~\cite{esp,
  mantovani_iccad20, carloni_dac16}. We enhanced the ESP cache hierarchy,
hardware monitors, and accelerator invocation API.

In summary, we make the following contributions:

\textbullet\ We classify the main accelerator cache-coherence modes for heterogeneous
  SoCs (Section~\ref{sec:background}).

\textbullet\ We make the case for the runtime selection of the cache-coherence mode of
  each accelerator (Section~\ref{sec:motivation}).

\textbullet\ We propose and evaluate \cohmeleon, a learning-based approach that
  transparently selects at runtime the best cache-coherence mode for each
  accelerator, without any prior knowledge of the target architecture
  (Sections~\ref{sec:reconfiguration},~\ref{sec:results}).

\textbullet\ The implementation of \cohmeleon, its FPGA-based experimental
infrastructure (Section~\ref{sec:evaluation}), and our enhancements to ESP are
released as part of the open-source ESP project\cite{esp}.

\section{Coherence Modes}
\label{sec:background}

\begin{table}
  \centering
\small 
	\begin{tabular}{@{}|l|>{\centering\arraybackslash}b{0.7cm}|>{\centering\arraybackslash}b{0.7cm}|>{\centering\arraybackslash}b{0.7cm}|>{\centering\arraybackslash}b{0.7cm}|@{}} 
    \toprule
    & {\bf non-coh DMA} & {\bf LLC-coh DMA} & {\bf coh DMA} & {\bf fully-coh} \\
    \midrule
Chen et al.~\cite{chen_iccd13} & \checkmark & & & \\
Cota et al.~\cite{cota_dac15}  & \checkmark & \checkmark & & \\
Fusion~\cite{fusion_isca15}    & & & \checkmark & \checkmark \\
gem5-aladdin~\cite{shao_micro16, bhardwaj_islped20, cong_aspdac21} & \checkmark & \checkmark & & \checkmark \\
Spandex~\cite{spandex_isca18} & & & & \checkmark \\
ESP~\cite{giri_ieeemicro18, giri_nocs18} & \checkmark & \checkmark & & \checkmark \\
NVDLA~\cite{nvdla} & \checkmark & & & \\
Buffets~\cite{buffets_asplos19} & \checkmark & & & \\
Kurth et al.~\cite{kurth_arxiv20} & \checkmark & & & \\
Cavalcante et al.~\cite{cavalcante_cf20} & & & \checkmark & \\
BiC~\cite{bic_dac11} & \checkmark & & & \\
Cohesion~\cite{cohesion_ieeemicro11} & & \checkmark & & \\
ARM ACE/ACE-Lite~\cite{arm_ace} & \checkmark & & \checkmark & \checkmark \\
Xilinx Zynq~\cite{nayak_ictis17, sadri2013energy, min_fpl19} & \checkmark & & \checkmark & \\
Power7+~\cite{power7_ibmrd13} & & & \checkmark & \\
Wirespeed~\cite{wirespeed_ibmrd10} & & & \checkmark & \\
Arteris Ncore~\cite{arteris_white_paper_2016} & & & \checkmark & \checkmark \\
CAPI~\cite{capi_ibmjrd15} & & & \checkmark & \\
OpenCAPI~\cite{opencapi16} & & & \checkmark & \\
CCIX~\cite{ccix17, ccix_intro} & & & \checkmark & \checkmark \\
Gen-Z~\cite{genz} & \checkmark & & & \\
CXL~\cite{cxl} & & & \checkmark & \checkmark\\
\bottomrule
  \end{tabular}
  \caption{Accelerator coherence modes in literature.}
  \label{tab:background}
  \vspace{-0.5cm}
\end{table}

The literature on cache coherence for accelerators has proposed several ways of
maintaining coherence in heterogeneous systems. These solutions range from
managing the coherence fully in hardware to managing it
fully in software, or with a hybrid of the
two methods\rev{~\cite{kelm_isca10,ace_arm11,leverich_taco08}}.

By extending existing classifications in literature~\cite{cota_dac15,
  giri_ieeemicro18, giri_nocs18, bhardwaj_cal19}, we identify four main types of
cache coherence for accelerators \cam{from a system perspective}.
We refer to these as \textit{accelerator cache-coherence modes}\cam{, which are
  defined independently from the specific cache-coherence protocol implemented
  by a given cache hierarchy}.
\rev{All modes \emph{always} keep data coherent, but they do so with different
  combinations of software-based and hardware-based coherence.}
The modes naming defines the degree of hardware coherence (non-coherent,
LLC-coherent, coherent), and at what level the accelerator accesses the memory
hierarchy: access to a private cache (cache access) or direct memory access on
the system interconnect (DMA).
\figurename~\ref{fig:soc-memory} depicts the accelerator interaction with the
memory hierarchy for each cache-coherence mode, while Table~\ref{tab:background}
lists the literature where these modes appear. \rev{The literature, including
  industry, has clearly not converged on a single coherence mode for LCAs. Each
  mode brings different benefits that can make it optimal depending on the
  situation.}

\textbf{Non-Coherent DMA.}
The accelerator does not have a private cache, and its memory requests bypass
the cache hierarchy and access main memory directly. In this approach, coherence
is implicitly managed by software.
If the accelerator data is allocated in a cacheable memory region, a flush of the
caches is required before invoking the accelerator to make sure main memory
contains the updated version of the data.
Some solutions allocate the accelerator data in virtual memory for an efficient
and transparent data sharing between processors and accelerators
(e.g. gem5-Aladdin~\cite{shao_micro16}).
Others reserve a contiguous physical memory region for the accelerator data to
avoid dealing with virtual address translation for the accelerator
(e.g. NVDLA~\cite{nvdla}).

\textbf{LLC-Coherent DMA.}
The accelerator does not have a private cache, and its memory requests are sent
directly to the LLC. The accelerator is kept coherent with the LLC, but not
with the private caches of the processor cores. Therefore, only the private
caches must be flushed before the accelerator execution.
Since the accelerator does not have a private cache, it normally sends memory
requests that have no notion of coherence, as, for example, when using the AMBA
AXI interface~\cite{arm_ace}. Then, the system around the accelerator enforces
the LLC-coherence.
For this reason, enabling the LLC-coherent mode requires a bridge to map
the non-coherent accelerator requests to the system's cache
protocol. LLC-coherent DMA has been demonstrated in literature on top of a MESI
cache hierarchy~\cite{sorin_slca11, giri_ieeemicro18, giri_nocs18, bhardwaj_islped20}.

\textbf{Coherent DMA.}
Similarly to LLC-coherent DMA, the accelerator does not have a private cache, and
its memory requests are sent directly to the LLC. However, in this case, the
cache hierarchy maintains full hardware coherence.
Therefore, no cache flush is needed, and the LLC recalls or invalidates data in
the private caches as needed.
Also in this case, the accelerators are normally non-coherent, requiring a
bridge to map their messages to the coherent system interconnect.
All the entries in Table~\ref{tab:background} with support for coherent DMA
have their own version of this bridge. For instance, ARM introduced the
accelerator coherency port (ACP) in the AXI protocol specification to allow the
connection of non-coherent accelerators to the coherent system
interconnect~\cite{arm_ace, ace_arm11}.
Coherent DMA, also referred to as \textit{I/O
  coherence}~\cite{arteris_white_paper_2016, cavalcante_cf20,cxl}, is a common
solution for cache-coherent chip-to-chip interconnects, as in the case of
connecting an accelerator to a CPU through a PCIe link~\cite{capi_ibmjrd15,
  opencapi16, ccix17}.

\textbf{Fully-coherent (Coherent Cache Access).}
A fully-coherent accelerator is equipped with a private cache, to which it sends
memory requests, instead of sending them directly on the system interconnect
like in the case of DMA. The coherence is handled fully in hardware, exactly
as for general-purpose processors.
There are various available cache-coherence protocols for the accelerator's
private cache. Beside standard protocols like MESI and MOESI~\cite{shao_micro16,
  giri_ieeemicro18, giri_nocs18}, other options are Fusion~\cite{fusion_isca15},
DeNovo~\cite{denovo_pact11}, or GPU-like coherence
protocols~\cite{nagarajan_slca20}.
Spandex acknowledges this diversity and introduces a flexible coherence
interface that enables systems with heterogeneous cache-coherence
protocols~\cite{spandex_isca18}.

\section{Motivation}
\label{sec:motivation}

Accelerators are heterogeneous by nature and this diversity is reflected in
their communication properties~\cite{lyons_taco12, cota_dac15}. They manifest a
wide range of memory access patterns, from long streaming bursts to single-word
irregular accesses. While an accelerator may read the same input data multiple
times, another may store intermediate results or write the output in place. Some
are highly computation-bound, requiring a very low communication
bandwidth. Others are communication-bound and benefit from an increased memory
bandwidth.
Furthermore, the same accelerator may show a wide range of communication properties
at runtime based on how it is configured when invoked;
some accelerators may even contain multiple engines, each optionally
selected at runtime.
Another important factor is the input size. If the accelerator's data is small
enough to fit in its local memory, it only needs to be loaded once from the
memory hierarchy; otherwise, many memory transactions are needed.
These varieties of communication properties and runtime variability contribute
to the complexity of optimizing the interactions between accelerators and the
memory hierarchy.

\begin{figure*}[t!]
\centering
\includegraphics[width=\linewidth]{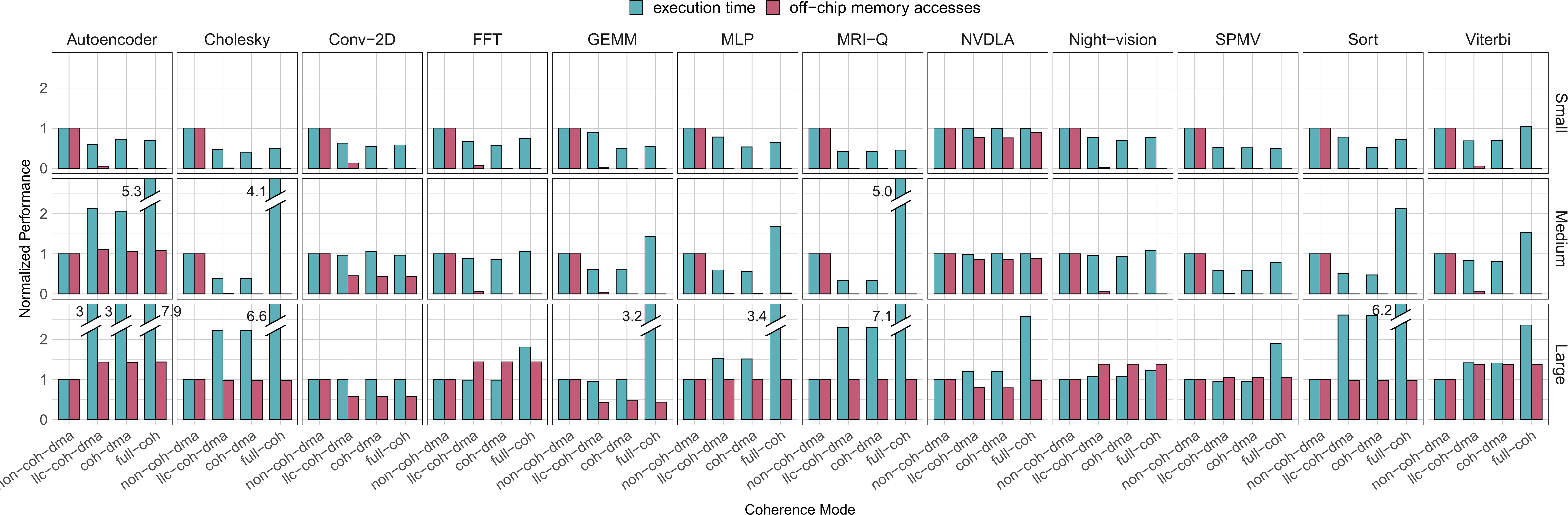}
\vspace{-0.7cm}
\caption{Accelerators running in isolation with different
coherence modes and workload sizes.}
\label{fig:motiv1}
\vspace{-0.5cm}
\end{figure*}

As mentioned in Section~\ref{sec:intro}, our work is focused on
\rev{fixed-function loosely-coupled} accelerators that execute coarse-grain
tasks. Normally, the synchronization between processors and accelerators and
across multiple accelerators does not occur at fine granularity (e.g. a cache
line), but rather at the level of an entire task, which corresponds to the
invocation of a particular accelerator. %
\rev{Even without fine-grained synchronization, the cache hierarchy can still be
  useful. The primary benefit is accessing an on-chip copy of the data (in case
  of a cache hit), as opposed to directly accessing off-chip memory. The caches can be
  especially useful when accelerators access the same data multiple times (e.g.,
  for storing and loading partial results), or when other components access data
  before or after the accelerator invocation (preparing the inputs or reading
  the outputs). Cache coherence also avoids the need for software-based
  synchronization mechanisms, such as cache flushing or copying accelerator data.
  The coherence mode determines whether the accelerators will leverage the
  caches to potentially mitigate the latency of accessing shared data. Thus, the
  selection of one of the four coherence modes dictates to which level of the
  cache hierarchy the accelerator memory requests should be routed to.}

\begin{table}
  \centering
\footnotesize
	\begin{tabular}{|p{0.18\columnwidth}|p{0.018\columnwidth}|p{0.018\columnwidth}|p{0.018\columnwidth}|p{0.018\columnwidth}|p{0.018\columnwidth}|p{0.018\columnwidth}|p{0.018\columnwidth}|p{0.018\columnwidth}|p{0.018\columnwidth}|p{0.018\columnwidth}|p{0.018\columnwidth}|p{0.018\columnwidth}|} 
          \toprule
    & \rotatebox[origin=c]{90}{Autoencoder}
    & \rotatebox[origin=c]{90}{Cholesky}
    & \rotatebox[origin=c]{90}{Conv2D}
    & \rotatebox[origin=c]{90}{FFT}
          & \rotatebox[origin=c]{90}{GEMM}
    & \rotatebox[origin=c]{90}{MLP}
    & \rotatebox[origin=c]{90}{MRI-Q}
    & \rotatebox[origin=c]{90}{\rev{NVDLA}}
    & \rotatebox[origin=c]{90}{Night-vision\footnotemark}
    & \rotatebox[origin=c]{90}{Sort}
    & \rotatebox[origin=c]{90}{SPMV}
    & \rotatebox[origin=c]{90}{Viterbi}
    \\
    \midrule
CortexSuite~\cite{cortexsuite} & & & \checkmark & & & \checkmark & & & & & & \\
ESP~\cite{mantovani_iccad20}   & \checkmark & \checkmark & \checkmark &\checkmark & \checkmark & \checkmark & \checkmark
                               & \checkmark & \checkmark & \checkmark & \checkmark & \checkmark \\
MachSuite~\cite{machsuite}     & & & & \checkmark & \checkmark & & & & & \checkmark & \checkmark & \checkmark \\
Parboil~\cite{parboil}         & & & & & \checkmark & & \checkmark & & \checkmark & \checkmark & & \\
PERFECT~\cite{perfect}         & & & \checkmark & \checkmark & & & & & \checkmark & & \checkmark & \\
S2CBench~\cite{s2cbench}       & & \checkmark  & & \checkmark & & \checkmark & & & & \checkmark & & \\
\bottomrule
        \end{tabular}
  \caption{The accelerators in this work target relevant applications that appear in various benchmark
    suites.}
  \label{tab:accelerators}
  \vspace{-0.5cm}
\end{table}
\footnotetext{The benchmark suites contain a subset of the night-vision pipeline.}
  
To establish the advantage of managing the accelerator-memory interactions
dynamically at runtime, we performed a series of experiments by leveraging the
ESP open-source SoC platform~\cite{mantovani_iccad20}. We designed a
many-accelerator multi-core SoC architecture and we evaluated it on FPGA.  We
used 11 accelerators available in the ESP release for the following tasks:
Denoising Autoencoder for the Street View House Numbers (SVHN) image dataset,
Cholesky decomposition, 2D convolution, 1D Fast Fourier Transform (FFT), dense
matrix multiplication (GEMM), multi-layer perceptron (MLP) classifier for the
SVHN dataset, magnetic resonance imaging (MRI-Q), \rev{the NVDLA~\cite{nvdla,
    giri_ieeemicro21}}, a ``night-vision'' application consisting of four
internal engines (noise filtering, histogram, histogram equalization, discrete
wavelet transform), sort, sparse matrix-vector multiplication (SPMV), and
Viterbi algorithm.  We built an SoC with 11 accelerators, one per type. These 11
accelerators target relevant applications amenable for hardware acceleration
that appear in a similar form in various benchmark suites, as shown in
Table~\ref{tab:accelerators}.
\rev{These accelerators are highly optimized; they employ custom local memory
  and memory access patterns, use a pipelined datapath that overlaps
  communication with computation, and exploit data reuse as much as
  possible. These accelerators are also flexible; they can be configured in
  different operating modes, to operate on batches of inputs, and with a wide
  range of input sizes}.
\rev{All the accelerators used in this work are designed with no notion of
  coherence. They merely send out memory requests, and the surrounding system
  transparently offers different ways, i.e. coherence modes, to handle these
  requests.}
These accelerators form a good assortment of fixed-function LCAs in terms of
memory access patterns.

\textbf{Accelerator Execution in Isolation.}
First, we evaluated the four cache-coherence modes described in
Section~\ref{sec:background} with each accelerator running alone and processing
three different workload data sizes: roughly 16KB (\textit{Small}), 256KB
(\textit{Medium}) and 4MB (\textit{Large}).  Each processor and accelerator has
its own 32KB private cache.  The 1MB LLC is split in two units, each
corresponding to a contiguous partition of the global address space and equipped
with a dedicated memory controller to access that partition.

\figurename~\ref{fig:motiv1} shows the results in terms of execution time (blue)
and off-chip memory accesses (red) for all the combinations of cache-coherence
mode and workload size.
Each bar shows the average of ten executions, normalized against the
non-coherent DMA results. These measurements include the overhead of invoking
the accelerator, (i.e., the execution of the accelerator's device driver and any
required cache flushes).

Since an application initializes all the data before invoking an accelerator,
the data is ``warm'' when the accelerator starts.  Hence, for {\em Small} and
{\em Medium} workloads, some coherence modes have zero off-chip accesses (i.e.
the red bar is missing), because all data is already loaded and fits in the
caches. Instead, the {\em Large} workloads are sized such that they do not fit
in the cache hierarchy, so data warm-up does not benefit performance.

The results in \figurename~\ref{fig:motiv1} show that the best cache-coherence
mode can vary at runtime based on the workload size and on the accelerator
type. In fact, given an accelerator, the winner is not the same for all
workloads sizes, and given a workload size, the winner is not the same for all
accelerators.
For instance, in the case of the {\em autoencoder}, the non-coherent mode goes
from being the slowest mode and by far the worst in terms of memory accesses
({\em Small} workload), to being at least three times faster than all the other
modes while incurring 30\% less memory accesses ({\em Large} workload). While a
few accelerators behave similarly, others, like GEMM, never have the
non-coherent mode as the best option.

The modes that do not require the device driver to flush the caches have a
smaller invocation overhead and may benefit from cached data. Hence, they tend
to perform best for smaller workload sizes \rev{(e.g. MLP Small/Medium)}.  The
non-coherent mode becomes more effective for larger workloads, which do not fit
in the caches and make the invocation overhead negligible.
\rev{For large workloads, in the case of some accelerators, the non-coherent
  DMA has fewer off-chip accesses than the other modes (e.g. FFT Large). In
  fact, for large workload sizes, the modes that use the caches can incur
  thrashing, i.e. the miss rate is very high and causes a large amount of cache
  evictions.}
As the workload size increases beyond 4MB, this phenomenon is bound to increase
in favor of the non-coherent mode, which bypasses the caches.

\rev{The non-coherent DMA mode always accesses off-chip data directly, but when
  the accelerator requests long bursts it can sustain higher throughput than the
  modes using the caches. Hence, it can have better performance even with more
  memory accesses (e.g. Cholesky Large).}

\begin{figure}[t!]
\centering
\includegraphics[width=\columnwidth]{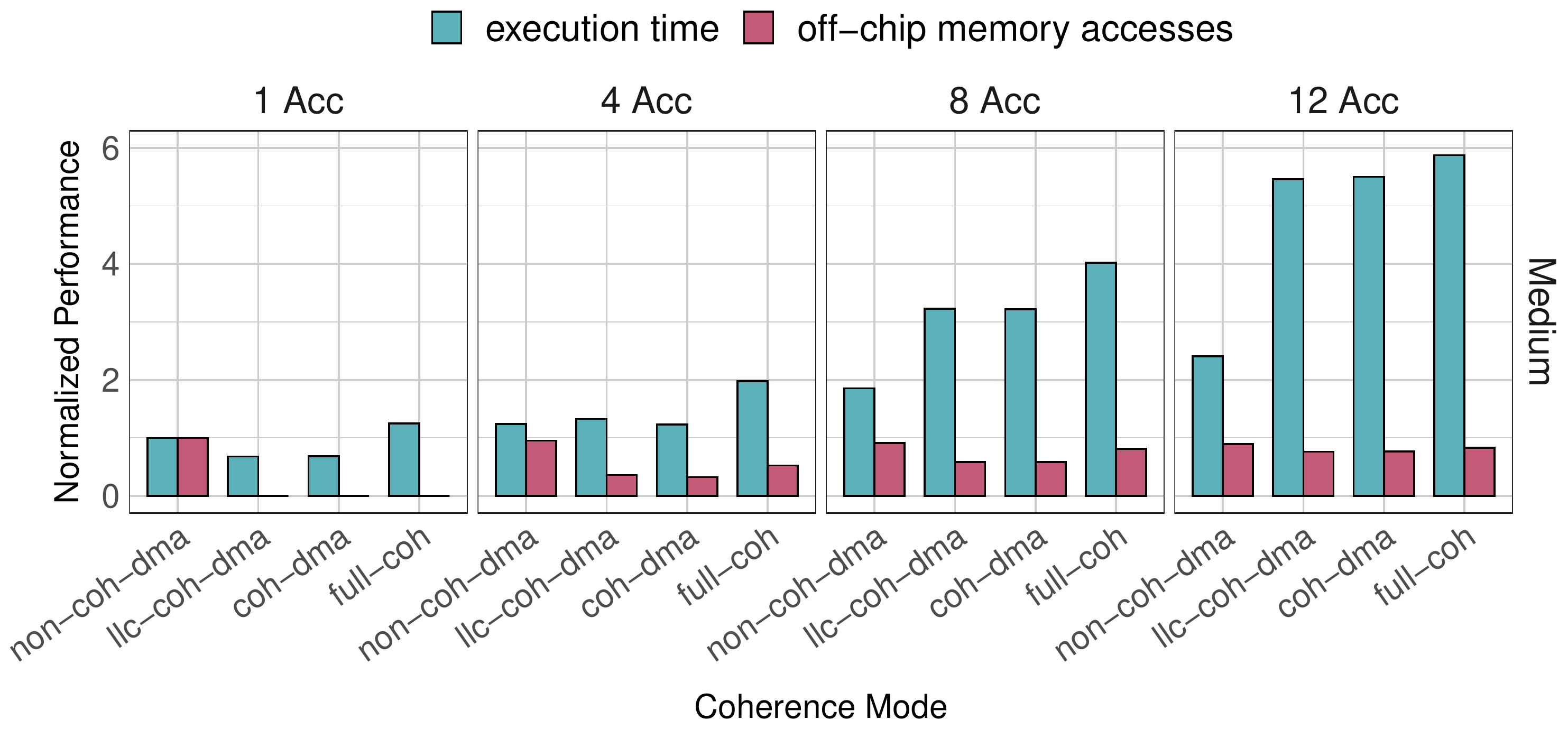}
\vspace{-0.7cm}
\caption{Accelerators running in parallel.}
\label{fig:motiv2}
\vspace{-0.5cm}
\end{figure}

\textbf{Parallel Accelerator Execution.}
Modern SoCs frequently contain up to tens of hardware accelerators, and
applications executing in parallel can cause multiple hardware accelerators to
run simultaneously. We thus study the performance degradation due to concurrent
execution of multiple accelerators and report the results of our study in
\figurename~\ref{fig:motiv2}. We chose medium-sized (256KB) workloads for each
of accelerator types and ran multiple experiments with 1, 4, 8 and 12
accelerators executing concurrently. We built an SoC with 12 accelerators, with
3 instances of FFT, Night-vision, Sort, and SPMV. Each accelerator is invoked
multiple times in a row from a separate software thread. For each set of
experiments, we averaged the performance of each accelerator over multiple
executions and we normalized it against the non-coherent DMA results for the
single accelerator execution. Then, we averaged the normalized performance of
all four accelerators to produce the results in \figurename~\ref{fig:motiv2}.

As the number of active accelerators increases, the non-coherent mode appears
to suffer the least, recording an execution time loss of up to $2.4\times$ with 12
accelerators, whereas the value of off-chip accesses stays constant.
The other modes, which could benefit from cached data in the case of
standalone accelerators, see an increase of memory accesses due to 
contention on the caches. 
Coherent DMA suffers the
worst slowdown: $8\times$ higher execution time when 12 accelerators run
concurrently compared to the single accelerator case.

\textbf{The Case for a Learning-Based Approach.}
The results presented in this section and summarized in
\figurename~\ref{fig:motiv1} and \figurename~\ref{fig:motiv2} highlight that
no fixed coherence solution is close to optimal for heterogeneous SoCs
integrating multiple types of \rev{fixed-function LCAs}.
Furthermore, these results suggest that deriving an analytic solution in the form of a
heuristic would require a tremendous effort to search a huge design space 
that consists of a large number of variables to consider.
For instance, the penalty of each
particular type of data transfer may differ depending on the type of coherent
interconnect of the system; the number of available accelerators may vary; the number of
concurrent accelerator invocations can change over time; the size of the
workload might not be fixed; updates to the system software could impact
scheduling priorities.

\cam{Our motivating results, collected with the implementation of the 4
  coherence modes provided by ESP, are aligned with the findings of the
  state-of-the-art literature presented in Table~\ref{tab:background}, which are
  based on a variety of different coherence modes implementations}.

These considerations led us to the development of \cohmeleon: \rev{the first}
learning-based approach to the orchestration of accelerator coherence in
heterogeneous SoCs.  As we describe the implementation of \cohmeleon in
Section~\ref{sec:reconfiguration} and evaluate its advantages in
Section~\ref{sec:results}, we show how the learning approach can eliminate the
need for a human-in-the-loop and avoid a lengthy design-space exploration that
would have to be repeated whenever some of the state-space variables change in
the system. Furthermore, our learning approach can optimize concurrently a
multi-objective reward function, which leads to improved performance with fewer
accesses to external memory.

\section{Cohmeleon}
\label{sec:reconfiguration}

We now present \cohmeleon, our solution for runtime orchestration of
accelerator coherence.  \cohmeleon's ability to adaptively
reconfigure the memory hierarchy of a heterogeneous SoC is built upon features
that span various levels of the hardware-software stack. 
We propose a general framework for runtime coherence selection, highlighting 
the necessary hardware and software requirements. Then, we show how reinforcement
learning can be applied to this framework to enable online and continuous learning.
Finally, we discuss the implementation of our approach in a particular SoC platform.

\vspace{-0.35cm}
\subsection{Runtime Reconfiguration}

Our general framework is divided into four main phases. During each accelerator
invocation we first \emph{sense} the conditions of the SoC using a minimal set
of status variables. This information is used to \emph{decide} which coherence
mode to apply to an accelerator invocation following a given policy. This
decision is immediately \emph{actuated}, and then its performance is
\emph{evaluated} using hardware monitors when the accelerator execution
completes.

\textbf{(1) Sense}:
	We aim at achieving system introspection with respect to
	\cam{accelerator performance under cache coherence modes} by making the system
	capable of continuously observing its own operation.  We propose a lightweight
	software layer to keep track of important system variables.  Since tracking the
	complete state of an SoC is intractable, we identified the following set of
	parameters in order to capture a compact ``snapshot" of the system while keeping
	the overhead of doing so small:
	
	\textbullet\ Number of active accelerators.

	\textbullet\ Coherence mode of each active accelerator.

	\textbullet\ Memory footprint (workload size) of active accelerators.

	In Section \ref{sec:motivation}, we saw that the number of active
	accelerators and the workload size affected
	performance.  Furthermore, the coherence and memory footprint of other
	active accelerators are relevant because the active coherence choices 
	can cause resource contention, and the total
	amount of active data impacts the efficacy of caches.

\textbf{(2) Decide}: The coherence decision follows a \textit{policy}, a set
	of rules that dictate the runtime decision given the current state.
	We consider various policies, including those 
	generated by our reinforcement learning module and several baseline
	policies.

\textbf{(3) Actuate}: We choose to configure an accelerator's
	coherence once per invocation, since this is a natural point of synchronization,
	and fixed-function accelerators tend to have uniform behavior throughout an
	invocation. In addition, selecting the coherence mode at invocation time
        incurs no overhead, because it happens concurrently to the
        application-specific configuration of the accelerator.

        How the actuation of coherence mode is performed depends on the underlying SoC
	implementation, but the important prerequisite is that the coherence mode can
	be changed at runtime. Changing the coherence mode requires hardware support
	for multiple policies, as opposed to a static choice of coherence mode for each
	accelerator or the entire SoC. \rev{\cohmeleon does not necessarily require support
	for all four coherence modes; it makes the selection based on the options that are
	available.} In general, support for particular
	coherence modes can be decoupled from the accelerator design by using
	mechanisms provided by the surrounding SoC. Hence, designers need
	not worry about coherence when creating accelerators for a system
	that utilizes \cohmeleon.

\textbf{(4) Evaluate}: To evaluate the quality of an accelerator's
	    invocation, we identify four metrics to observe:
 
	    \textbullet\ Total execution time, including any overhead due to
		    accelerator invocation (i.e device driver, cache flushes).

	    \textbullet\ Off-chip memory accesses during the invocation. 

	    \textbullet\ Total cycles in which the accelerator is actively executing.

	    \textbullet\ Total cycles in which the accelerator is communicating
		    (issuing a request or awaiting a response) with memory.
 
	Total execution time and off-chip memory accesses are obvious
          performance attributes. We choose the other two metrics to account for
          compute-bound accelerators. For such accelerators, the total execution
          time may not change even if the memory system performs better, but the
          ratio of communication cycles to total cycles would get smaller.

	While execution time can be measured in software, the remaining values
	must be measured in hardware. Hence, we propose the addition of
	a lightweight hardware monitoring system that can be integrated easily
	into any SoC.
        Our monitoring system continuously exposes these values to software, thus
        allowing for the evaluation of an accelerator's performance to inform
        an intelligent selection of coherence modes.

\vspace{-0.35cm}
\subsection{Reinforcement Learning Module}
\label{sec:rl}

In reinforcement learning (RL), an autonomous agent learns behaviors through 
trial-and-error interactions with the environment~\cite{Kaelbling1996, 
Sutton1998}. At each step, the agent perceives the {\em state} (sense), chooses 
an {\em action} (decide), takes the action (actuate), and receives a {\em 
reward} at the beginning of the next step (evaluate). The objective is  
to find an optimal {\em policy} that determines which action to choose 
at each state so that the expected reward is maximized.

For the runtime reconfiguration problem of \cohmeleon, we propose a 
variant of Q-learning,
a widely adopted RL
approach that does not require any model of the 
environment~\cite{Watkins1989, Watkins1992}. This approach has a number of 
advantages for the target problem. First, it enables an automatic discovery of 
a coherence-selection policy during regular SoC operation.
Second, it requires neither human expertise of the target 
SoC architecture nor offline-training data. Third, 
RL allows continuous updating of the coherence-selection 
policy. As an SoC runs new workloads and encounters new system states, 
a RL module can update its policy to improve the 
performance. Finally, we can optimize the system with respect to 
multiple objectives
by formulating the learning {\em reward} accordingly.

\begin{table*}[t]
\small
\centering
\setlength{\tabcolsep}{0.4em}
	\begin{tabular}{@{}m{0.85in}|m{4.485in}|m{1.4in}@{}} 
     \toprule
	 \textbf{\begin{tabular}[c]{@{}c@{}}State Attribute \end{tabular}}&
		\textbf{\begin{tabular}[c]{@{}c@{}}Description \end{tabular}}&
		\textbf{\begin{tabular}[c]{@{}c@{}}Values \end{tabular}} \\

     \midrule
 Fully coh acc & Total number of active fully coherent accelerators & 0, 1, 2+ \\
\midrule
 Non coh acc per tile & Avg no. of non-coh accelerators communicating with each memory partition needed by the target accelerator invocation & 0, 1, 2+ \\
\midrule
 To LLC per tile & Avg no. of accelerators accessing each LLC partition needed by the target accelerator invocation& 0, 1, 2+ \\
\midrule
 Tile footprint & Avg utilization of each partition of the cache hierarchy needed by the target accelerator invocation& $\leq$L2, $\leq$LLC Slice, $>$ LLC Slice \\
\midrule
 Acc footprint & Memory footprint of the target accelerator invocation & $\leq$L2, $\leq$LLC Slice, $>$ LLC Slice \\
\bottomrule
\end{tabular}
\caption{State space $S$: a state $s\in S$ is a 5-tuple of the following attributes.}
\label{tab:state_coh}
\vspace{-0.3cm}
\end{table*}

A Q-learning agent maintains a {\em Q-table} that stores, for each 
state-action pair, a {\em Q-value} that represents the expected reward of 
taking that action from that state. The agent can use many strategies to select 
an action from each state, including an \textit{epsilon-greedy} approach. 
At each step, this approach selects either a random action, with probability 
$\epsilon$, or the action with the highest Q-value based on the current state.
This encourages both \textit{exploration} of the action space and
\textit{exploitation} of the knowledge gained from prior trials. After each
action is taken, the reward is evaluated at the next step, and the Q-value
corresponding to that state-action pair is updated from the previous value with
a \textit{learning rate} $\alpha$.

\begin{figure}[t]
\centering
\includegraphics[width=0.9\columnwidth]{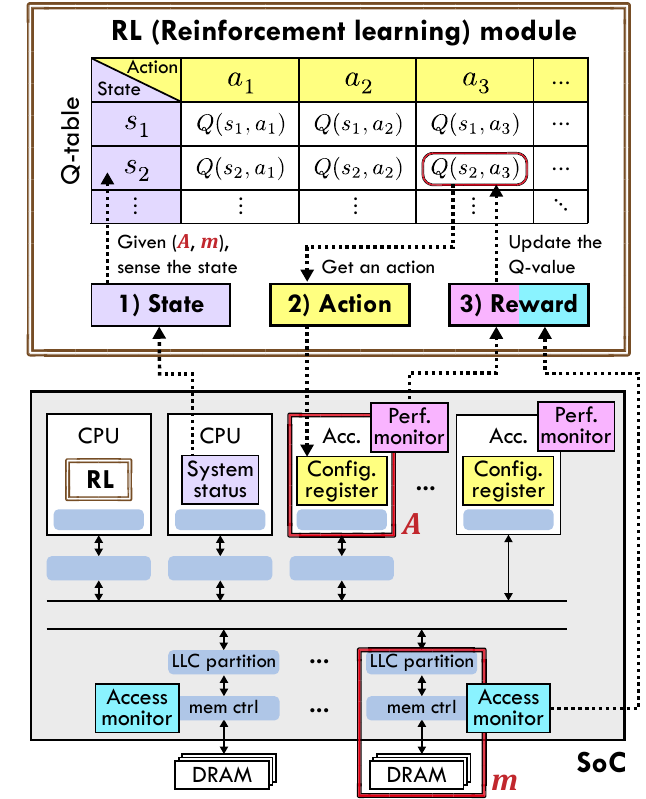}
\caption{Overview of the proposed learning module.}
\vspace{-0.5cm}
\label{fig:q-learning}
\end{figure}

\textbf{Model definition.}
\figurename~\ref{fig:q-learning} shows an overview of the proposed learning 
module for coherence selection. The problem is modeled with the following 
states, actions, and rewards.

\textbf{(1) States:}
\rev{Based on the results in the motivation section and on prior
  work~\cite{giri_aspdac19, bhardwaj_cal19},
we define} the state space $S$ with the following 5 attributes: {\em
Fully-Coherent-Acc, Non-Coh-Acc-per-Tile, To-LLC-per-Tile, Tile-Footprint,} and
{\em Acc-Footprint}. Each attribute can have one of three possible values, as
described in Table~\ref{tab:state_coh}. A state $s\in S$ is a 5-tuple of these
attributes, and hence, $|S| = 3^5 = 243$.
The value of the state $s$ for any accelerator invocation is used 
to index the Q-table of \figurename~\ref{fig:q-learning}. 

\textbf{(2) Actions:}
The action set $A$ contains the 4 coherence modes: {\em non-coherent, 
LLC-coherent-DMA, coherent-DMA,} and {\em fully-coherent}. Thus, the 
coherence Q-table contains $|S|\times|A|=243\times4=972$ entries. The action
step sets the coherence configuration register of the given accelerator tile.

\textbf{(3) Rewards:}
As shown in \figurename~\ref{fig:q-learning}, when an accelerator is to be
invoked, the RL module senses the state $s$ and looks up the
Q-table to determine the best action (or, randomly chooses an action) $a$.
Then, the accelerator is invoked with the selected action. After it completes
execution, its reward is computed based on the performance of both the
accelerator invocation and
the memory system during the invocation. 
To define the reward function, we
define three measurements of an 
invocation. For the $i$-th invocation of an accelerator $k$:

	\textbullet\ $exec(k, i)$ is the \textit{scaled execution time} - the
		total execution time divided by the footprint of the invocation.

	\textbullet\ $comm(k, i)$ is the \textit{communication ratio} - the number of accelerator
		communication cycles divided by the total number of execution cycles.

	\textbullet\ $mem(k, i)$ is the \textit{scaled off-chip memory access count} during the invocation - 
		the total number of memory accesses divided by the footprint of the computation.

Then, we define the three component functions:
\begin{align*}
	&R_{exec}(k, i) = \frac{\min_{j \le i} \left[exec(k, j)\right]}{exec(k, i)} \\
	&R_{comm}(k, i) = \frac{\min_{j \le i} \left[comm(k , j)\right]}{comm(k, i)} \\
	&R_{mem}(k, i) = 1 - \frac{mem(k, i) - \min_{j \le i} \left[mem(k, j)\right]}{\max_{j \le i} \left[ mem(k, j) \right] - \min_{j \le i} \left[ mem(k, j) \right]}
\end{align*}

For the
$exec$ component, we store the best scaled execution time for each accelerator
thus far. We can see that smaller execution times maximizes the ratio in
$R_{exec}$. As previously mentioned, we utilize the $comm$ part
to account for compute-bound accelerators. If the memory system performs
better, this ratio will be lower. Again, we see that $R_{comm}$ is maximized
for smaller values of the communication ratio. $R_{mem}$ takes a different
form because accelerator invocations may cause zero off-chip memory accesses.
Using both the maximum and minimum scaled access counts, the
presented equation maps higher memory-access counts near zero and lower
counts near one. 
Finally, the reward $R(s, a; k, i)$ for the $i$-th invocation of
accelerator $k$ with action $a$ from state $s$ is 
$$R(s, a; k, i) = x \cdot R_{exec}(k,i) + y \cdot R_{comm}(k,i) + z \cdot R_{mem}(k,i)$$

where $x$, $y$, and $z$ are constant weights that can be tuned.

\textbf{Training.}
At the beginning of training, all entries in the Q-table are set to zero.
The table is updated in the following manner. Whenever
an accelerator is invoked, the state is recorded.
After the accelerator completes execution, the reward is computed and is used to
update the Q-table for the recorded state and chosen action, as follows:
$$Q(s, a) \leftarrow (1 - \alpha) \times Q(s, a) + \alpha \times R(s, a)$$

where $R(s, a)$ is the reward that results from taking action $a$ out of state
$s$, and $\alpha$ is the learning rate.

\vspace{-0.2cm}
\subsection{Implementation}
\label{sec:fpga}

\rev{As shown in Table~\ref{tab:background}, there is already a number of
  architectures in literature that support multiple coherence modes for
  accelerators (e.g. ARM ACE/ACE-Lite, gem5-aladdin) and are amenable for
  hosting \cohmeleon.}
We implemented and evaluated \cohmeleon as part of a comprehensive FPGA-based
infrastructure that we developed to study accelerator coherence-mode selections
in many-accelerator SoCs. Our implementation is based on ESP, an open-source
platform for agile SoC design~\cite{mantovani_iccad20}.
\rev{We chose ESP because it supports the runtime selection of three of the
coherence modes we identified in Section~\ref{sec:background} and has a suite of
fixed-function LCAs. Furthermore, ESP allows for rapid prototyping of
many-accelerator SoCs on FPGA and is open source.}

The ESP SoC platform automates the integration of processor cores and
accelerators into a grid of tiles connected by a 2D-mesh multi-plane NoC.
\rev{There are four main types of tiles: processor tile, accelerator tile,
  memory tile for the communication with main memory, and auxiliary tile for
  peripherals (e.g. UART or Ethernet) or system utilities (e.g. the interrupt
  controller). In this work the processor tile contains the SPARC 32-bit LEON3
  processor core~\cite{leon3} from Cobham Gaisler. All components are connected
  by ESP's network-on-chip, which counts 6 32-bit physical planes, with one
  cycle latency between neighboring routers. Each memory tile has a link to main
  memory with bandwidth of 32 bits per cycle.}
The ESP cache hierarchy matches the one represented in \figurename~\ref{fig:soc-memory} and
is distributed across processor tiles, which include an L2 
private cache, and memory tiles. Each memory tile hosts a partition of the LLC,
a DRAM controller, and a dedicated channel to the corresponding partition of
off-chip main memory.
In addition, the accelerator tile can optionally integrate a private cache to enable
the fully-coherent mode. 
\rev{With the exception of the optional private cache, ESP's support for the
runtime selection from multiple coherence modes adds negligible area to the
SoC.} Furthermore, coherence is handled by the ``socket" surrounding the
accelerator, so the accelerators are designed with no notion of coherence.

\textbf{(1) Sense}:
We implemented our introspective SoC status tracking in the ESP accelerator
invocation API, which is a set of user-space functions for invoking loosely-coupled
accelerators from software applications \cite{mantovani_iccad20}. We defined new
global structures containing the number of active accelerators, their
footprints, and the chosen coherence mode. When an accelerator is invoked -- and
when it returns control to software -- this structure is updated accordingly.

\textbf{(2) Decide}:
The decision-making for a runtime coherence \textit{policy}
is also implemented in the back-end of the ESP API.
Here, we outline several possible policies to compare to our RL approach.  The
\textit{Random} policy randomly chooses a coherence mode for each accelerator
invocation at runtime.  The \textit{Fixed} policy statically selects the same
coherence mode for each accelerator invocation, mimicking a design-time
decision. \rev{This represents nearly all previous work.} The \textit{Fixed}
policy can either be \textit{homogeneous}, where every accelerator operates with
the same coherence mode, or \textit{heterogeneous}, where the coherence mode can
be chosen independently for each accelerator. In the heterogeneous case, we
choose a coherence mode based on profiling the accelerator's performance in each
mode while sweeping the footprint of the workload on different invocations.
\rev{The \textit{fixed-heterogeneous} policy is used as a comparison to prior
  design-time solutions that select a fixed coherence mode for each accelerator
  \cite{bhardwaj_cal19, bhardwaj_islped20}}.

Next, we present an introspective, manually-tuned algorithm, that chooses the
coherence mode based on the status of the system. We designed this algorithm to
minimize the runtime for accelerators in an ESP SoC. We used data from tens
of thousands of accelerator invocations, combined with knowledge of ESP's
implementation of the coherence modes, to produce a highly optimized policy,
shown as Algorithm~\ref{alg:manual}. \rev{The \textit{manual} algorithm is used
as a comparison to prior approaches that use static heuristics to select
a coherence mode at runtime \cite{giri_aspdac19}}.
If applied to SoC architectures other than our target (i.e. ESP) this algorithm
would need manual tuning and possibly some major adjustments. For instance, an
SoC that uses a more-optimized coherence protocol for accelerators than MESI
could benefit from an increased reliance on the fully-coherent mode, but this
manual algorithm would not select it more frequently.

In contrast, the RL module we presented in Section
\ref{sec:rl} generates its own coherence-selection policy. In
learning epochs where the agent ``chooses" to explore the state space, it
follows a random policy. When attempting to maximize the reward, however, the
model selects the coherence mode with the highest Q-value from the current
state. The Q-table thus dictates the coherence decision given the current state,
but unlike other policies, it is \emph{adaptive} and can change with the
addition of new information.

\newlength{\textfloatsepsave} \setlength{\textfloatsepsave}{\textfloatsep}
\setlength{\textfloatsep}{0pt}
\begin{algorithm}[t]
  \Small
  \caption{Manually-tuned coherence mode selection.}
    \label{alg:manual}
\begin{algorithmic}
	\IF{$\text{footprint} \leq \text{EXTRA\_SMALL\_THRESHOLD}$}
        \STATE $\text{coh} \leftarrow \text{FULLY-COH}$
        \ELSIF{$\text{footprint} \leq \text{CACHE\_L2\_SIZE}$}
        \IF{$\text{active\_coh\_dma} > \text{active\_fully\_coh}$}
        \STATE $\text{coh} \leftarrow \text{FULLY-COH}$
        \ELSE
        \STATE $\text{coh} \leftarrow \text{COH-DMA}$
        \ENDIF
        \ELSIF{$\text{footprint} + \text{active\_footprint} > \text{CACHE\_LLC\_SIZE}$}
        \STATE $\text{coh} \leftarrow \text{NON-COH}$
        \ELSE
        \IF{$\text{active\_non\_coh} \geq 2$}
        \STATE $\text{coh} \leftarrow \text{LLC-COH-DMA}$
        \ELSE 
        \STATE $\text{coh} \leftarrow \text{COH-DMA}$
        \ENDIF
        \ENDIF
\end{algorithmic}
\end{algorithm}

\textbf{(3) Actuate}:
Each coherence decision is actuated by the \cam{accelerator} device driver's writing to a
memory-mapped register in the accelerator tile that controls the mechanism by
which the accelerator communicates with memory. Because accelerators share
cacheable memory with processors in ESP, the LLC-coherent-DMA and
non-coherent-DMA modes require software-managed cache flushes, as described in
Section \ref{sec:background}, before the accelerator can begin executing.
\rev{It is possible to handle the flushes in a way that is completely
  transparent to} \rev{the programmer, which is the case in ESP.
In fact, \cohmeleon actuates the coherence mode with a single line of code.}
Using a custom MESI directory-based and NoC-based protocol, ESP supports
runtime selection for all of the coherence policies, excluding coherent-DMA.
We extended the protocol to support coherent-DMA by issuing recalls from the
LLC to a private cache when the private cache holds data that is the target of
a DMA request. By adding support for the coherent-DMA model, we did not
introduce any area overhead in the accelerator tiles. 

 \setlength{\textfloatsep}{\textfloatsepsave}

\textbf{(4) Evaluate}:
We developed a new hardware monitoring system to 
measure our chosen metrics of accelerator performance. In each tile, we added a set of
memory-mapped registers to a pre-existing APB interface. Each register
is connected to logic that increments its value when the corresponding
phenomenon occurs. These monitors are distributed across all tiles but are accessible from
software through a single contiguous region of the I/O address space using
standard Linux system calls, such as \texttt{\small mmap} and \texttt{\small ioremap}.

We access the hardware monitors from the accelerator device driver. The
accelerator cycle counters, which are reset at the beginning of its execution,
are read at the end of the invocation. Memory access counters are read before and
after each invocation to determine the change, potentially accounting for
overflow. We note that when multiple accelerators are communicating with
a memory controller, we cannot know the exact number of memory accesses
caused by each accelerator without additional hardware support to track
which accelerator's transactions cause misses or evictions in the LLC. Instead, in
order to minimize \cohmeleon's required hardware support, we
chose to approximate the number of off-chip memory accesses for a
particular accelerator. Our approximation relies on the assumption that larger
workloads will, in general, trigger more memory accesses. This
works particularly well for the non-coherent mode, since all data must be
brought in from off chip, and the other modes when running workloads that do
not fit in the caches. Furthermore, more transactions likely mean more cache
misses and evictions.  We thus define the memory accesses caused by accelerator
$k$ at a memory controller $m$ as:
$$ddr(k, m) = ddr\_total(m) \times \frac{footprint(k, m)}{\sum_{acc \in A}
footprint(acc, m)}$$

where $A$ is the set of active accelerators, $ddr\_total(m)$ is the observed
change in off-chip memory accesses at memory controller $m$, and $footprint(acc,
m)$ is the memory footprint of accelerator $acc$ allocated to memory controller
$m$. By this definition, each accelerator's share of the total memory accesses
will be proportional to its active memory footprint. Better approximations are
likely possible with some knowledge of accelerator characteristics, but we
chose this approximation to keep \cohmeleon accelerator-agnostic.

\section{Evaluation Strategy}
\label{sec:evaluation}

\begin{table}
\footnotesize
\centering
	\begin{tabular}{p{0.130\columnwidth}|p{0.08\columnwidth}|p{0.08\columnwidth}|p{0.08\columnwidth}|p{0.08\columnwidth}|p{0.08\columnwidth}|p{0.08\columnwidth}|p{0.08\columnwidth}} 
          \toprule
& \multicolumn{4}{p{0.32\columnwidth}|}{SoCs w/ Traffic Generator} & \multicolumn{3}{p{0.24\columnwidth}}{Case Study SoCs} \\
& \textbf{SoC0} & \textbf{SoC1} & \textbf{SoC2} & \textbf{SoC3} & \textbf{SoC4} & \textbf{SoC5} & \textbf{SoC6} \\

     \midrule

	 Accelerators &  12  &  7  &  9 &  16 &  11 & 8 & 9 \\ 
	 NoC size &  5x5 &  4x4 &  4x4 &  5x5 &  5x4 & 4x4 & 4x4 \\
	 CPUs &  4 &  2 &  4 &  4 &  2 & 1 & 1 \\ 
	 DDRs &   4 &  4  &  2 &  4 &  4 & 4 & 2\\ 
	 LLC part. &  512kB &  256kB &  512kB & 256kB &  256kB & 256kB & 256kB \\ 
	 Total LLC &   2MB  &   1MB &  1MB &  1MB &  1MB & 1MB & 512kB \\ 
	 L2 cache &  64kB &  32kB &  32kB  &  64kB &  32kB & 32kB & 32kB \\ 
     \bottomrule
        \end{tabular}
  \caption{Parameters of the evaluation SoCs.\label{tab:synthsocs}}
\vspace{-0.7cm}
\end{table}

To evaluate \cohmeleon, we realized multiple SoC prototypes and test
applications that serve as full-system case studies.

\textbf{Traffic-Generator.}
From the viewpoint of the rest of the SoC, an accelerator can be characterized
by its patterns of communication with the memory hierarchy. After analyzing many
accelerators, we derived a list of basic accelerator properties that influence
these patterns. Then, we designed a traffic-generator that is
configurable with respect to these properties, allowing us to efficiently study
the diverse set of communication patterns that are expressed by
accelerators. The parameters of the traffic-generator are access pattern
(streaming, strided, or irregular), DMA burst length, compute duration, data
reuse factor, read-to-write ratio, stride length (for strided patterns), access
fraction (for irregular patterns), and in-place storage.

\textbf{SoCs.}
We implemented FPGA-based prototypes of four different SoCs utilizing the
traffic-generator, and three Case Study SoCs utilizing the same set of
accelerators from Section~\ref{sec:motivation}.
For these designs, Table \ref{tab:synthsocs} reports the key parameters, which
we vary in order to validate that our contributions can generalize to different SoC
configurations. All accelerators are equipped with a private cache, except five
of the accelerators of SoC3, \cam{which could not be included due to resource
constraints of the chosen FPGA.}

SoC4 has one instance of each of the 11 accelerators of
Table~\ref{tab:accelerators}, thus representing the modern SoC trend of invoking
many heterogeneous accelerators while running multiple applications in parallel.

SoC5 targets the domain of collaborative autonomous vehicles by embedding key
accelerators~\cite{sisbot_grcon19}. Two FFT and two Viterbi
accelerators support encoding and decoding for vehicle-to-vehicle (V2V)
communication. Two 2D Convolution (Conv-2D) and two Matrix Multiplication (GEMM)
accelerators support inference for convolutional neural networks (CNNs) for
object recognition and labeling. For this purpose the Conv-2D and GEMM
accelerators embed bias addition, pooling and activation capabilities.

SoC6 targets the computer vision domain by providing \rev{three instances of} an
image classification pipeline composed of three accelerators~\cite{giri_date20}:
night-vision, which has a 4-stage pipeline for processing dark images,
autoencoder for denoising images, and MLP for image classification.

\renewcommand{\tabcolsep}{3pt}

\begin{figure}[t]
\includegraphics[width=\linewidth]{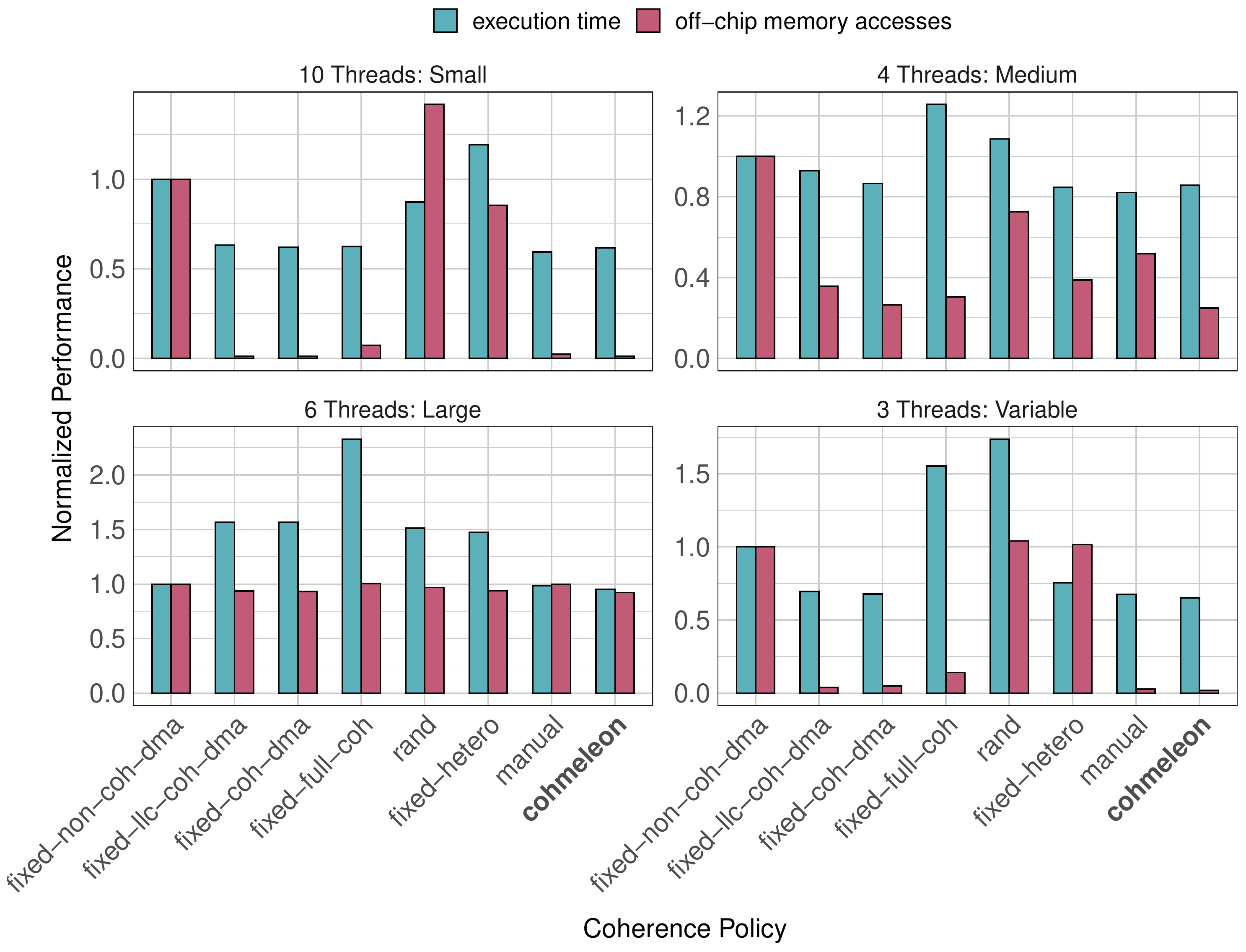}
\centering
\vspace{-0.7cm}
\caption{Performance of four phases of the evaluation application, varying
	threads count and workload sizes.} 
\label{fig:phases}
\vspace{-0.5cm}
\end{figure}

\textbf{Applications.}
We developed a multithreaded application to invoke the traffic-generator
in many different ways. The application consists of a set of
\textit{phases} that are each meant to represent a real application. A phase
consists of a number of threads. A thread consists of a single dataset and a
``chain" of accelerators, configured with different parameters, that operate
serially on that dataset - the output of one accelerator is the input to the
next. Optionally, each thread can loop over the accelerator invocations.
The application phases and parameters are specified using a configuration
file. Through our experiments, we ensured that the instances of the
application vary in terms of the number of accelerator threads running in
parallel, the workload sizes in use, and the configuration parameters of each
traffic-generator.

Furthermore, we developed multithreaded evaluation applications specific to each
Case Study SoC, with a structure similar to the applications for the
traffic-generators SoCs: organized in phases and designed to
stress multiple operating modes and workload sizes for each accelerator. For
each SoC, the application invokes pipelines of accelerators as appropriate for
the target domain. For example, the night-vision, autoencoder and MLP work in a
chain to undarken, denoise, and then classify the input images.
On SoCs containing multiple copies of the same accelerators, our evaluation
applications can parallelize the workload to improve the throughput.
For these applications, we define the following characterization of
workload sizes:
\emph{Small (S)} when smaller than accelerator's L2 cache;
\emph{Medium (M)} if smaller than one partition of the LLC;
\emph{Large (L)} when smaller than the aggregate LLC;
\emph{Extra-Large (XL)} if larger than the LLC.

\rev{We use ESP's solution for the allocation of accelerator
  data~\cite{mantovani_cases16}. Data is allocated in big Linux pages, so that
  it results in a relatively small page table that can normally fit in the
  accelerator tile's TLB. The TLB is loaded at the beginning of the accelerator
  invocation and it provides a miss-free address translation. This solution
  allows for true data sharing between CPUs and accelerators with no need for
  data copies or contiguous physical allocations. The overhead of loading the
  TLB and address translation is included in all results.}

\textbf{Experimental Setup.}
\rev{We deployed the full SoCs on a Xilinx Ultrascale XCVU440 FPGA. The LEON3
  cores (soft-cores) in the SoC run Linux SMP, on top of which we executed the
  evaluation applications for each of the above SoCs.}

\cohmeleon learns online at runtime, i.e. there is no offline training. For
each SoC we run a randomly configured instance of the evaluation application. The learning
coefficients are initialized to $\epsilon=0.5$ and $\alpha=0.25$ and decay
linearly to 0 over a selected number of iterations. Once the learning model has
converged, we disable further updates and evaluate its performance on a
different instance of the application.

To compare performance across policies we measured the total execution time and
off-chip memory accesses for each phase of the applications, including any
overhead of the accelerator invocations, such as cache flushes. Because the
phases vary in terms of the number of accelerators working in parallel and
workload sizes, we can compare how the different policies perform in many situations.
\vspace{-0.5cm}

\vspace{-0.25cm}
\section{Experimental Results}
\label{sec:results}

\textbf{Phase Analysis.}  We first present the results of four selected phases
of the evaluation application running on SoC0 (\figurename~\ref{fig:phases}).
These phases differ in the number of threads and workload sizes in order to
highlight both the variation of performance in policies and the coverage that
our evaluation applications provides. All results are normalized to the values
of the \emph{Fixed non-coherent DMA} policy.

\begin{figure}[t]
\includegraphics[width=\linewidth]{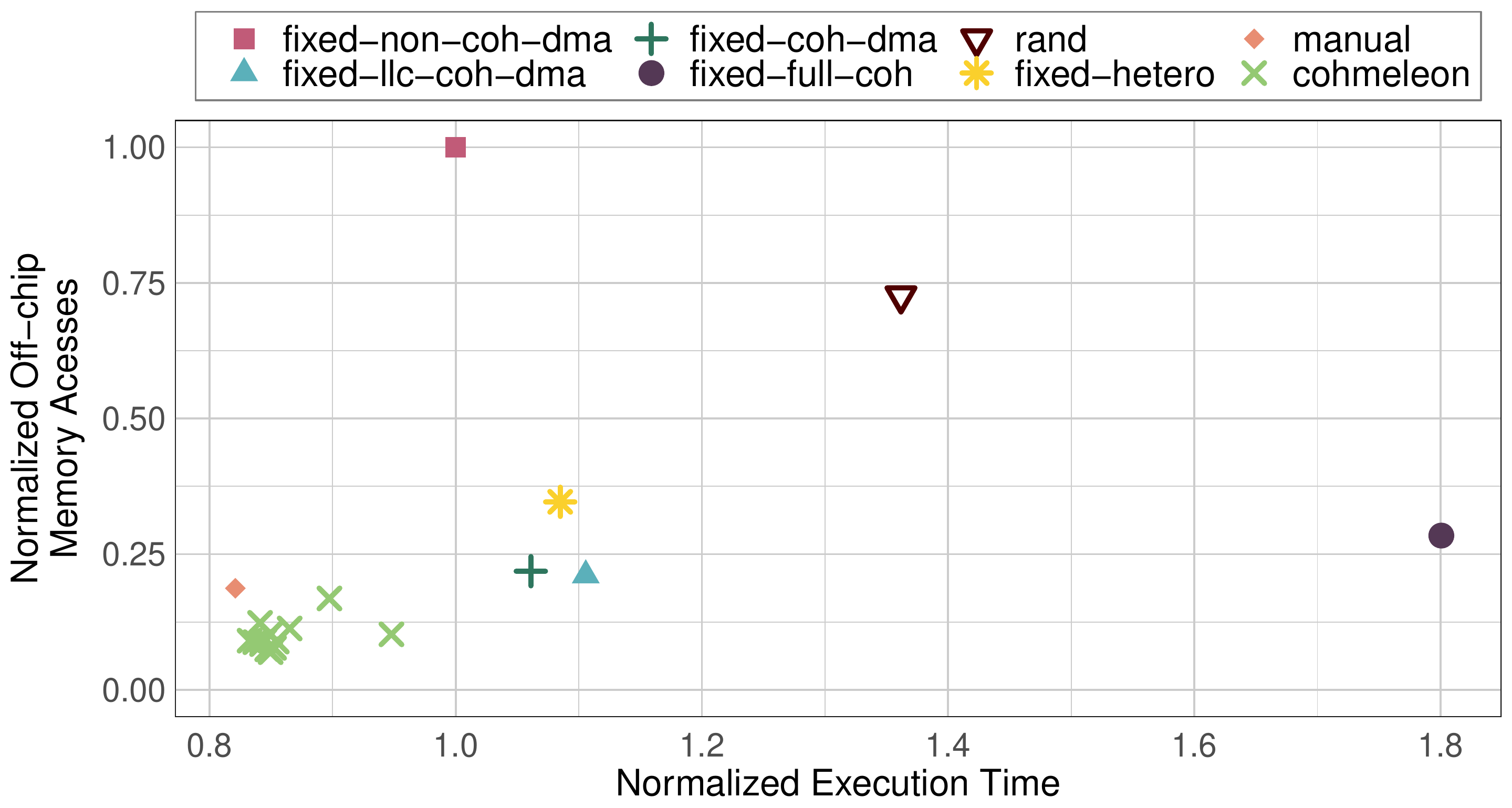}
\centering
\vspace{-0.6cm}
\caption{The impact of the reward function on SoC0.}
\label{fig:soc0_coh}
\vspace{-0.5cm}
\end{figure}

As the workload size \cam{and number of threads vary}, we observe variations in
the relative performance of the \textit{Fixed homogeneous} policies,
consistent with the motivation results in Section \ref{sec:motivation}. On the
other hand, both the manually-tuned algorithm and \cohmeleon match or improve
upon the best execution time of the other policies in all phases. Meanwhile, the
\textit{Fixed heterogeneous} policy is unable to achieve this result due to the
variation in workload size and system status. The manually-tuned algorithm
achieves similar execution times as \cohmeleon in all phases, but \cohmeleon
achieves that performance with fewer off-chip memory accesses.

\textbf{Design-Space Exploration of Reward Function.}
Next, we experiment with different reward functions on SoC0, only varying
  the values of the three $x, y, z$ coefficients of $R_{exec}$, $R_{comm}$, and
  $R_{mem}$.
We train \cohmeleon for 50 iterations of the evaluation application with each
reward function, and then test each model, as well as the other baseline
policies, on a different instance of the application.  For each policy, we
normalize the performance of each phase to the \emph{Fixed non-coherent DMA}
policy. In \figurename~\ref{fig:soc0_coh}, we plot the geometric mean of the
normalized execution time and off-chip memory accesses over all phases.
First, we highlight the large cluster of \cohmeleon data points in the bottom
left of the chart. \cohmeleon is capable of matching the
execution time of the manually-tuned algorithm, while achieving
the best value for off-chip memory accesses.  \cohmeleon's flexibility
in optimizing for multiple objectives clearly allows for the discovery of
policies that are near-optimal over multiple performance metrics.  However, we
notice that while \cohmeleon generates \textit{Pareto-optimal} data points, the
cluster of points does not present much variation. Thus, we cannot trade off an
improvement in execution time with a reduction of off-chip memory accesses or
vice versa. Indeed, off-chip memory accesses contribute to a relevant part of
the execution time of communication-bound accelerators.

We trained 15 different models, and most perform quite similarly. In fact,
only two points, which correspond to reward functions that weighed heavily
($>90\%$) for off-chip memory accesses, have significantly worse performance.
The remaining points have extensive variation in the reward function. For
instance, two of the Pareto-optimal points use reward functions that
give the following percentage weights to execution time, communication ratio, and off-chip
memory accesses, respectively: (a) 67.5, 7.5, 25 and (b) 12.5, 12.5, 75.  We
conclude that, for this particular architecture, near-optimal performance can be
achieved with a wide variety of reward functions without significant
trial-and-error.

\begin{figure}[t]
\includegraphics[width=\linewidth]{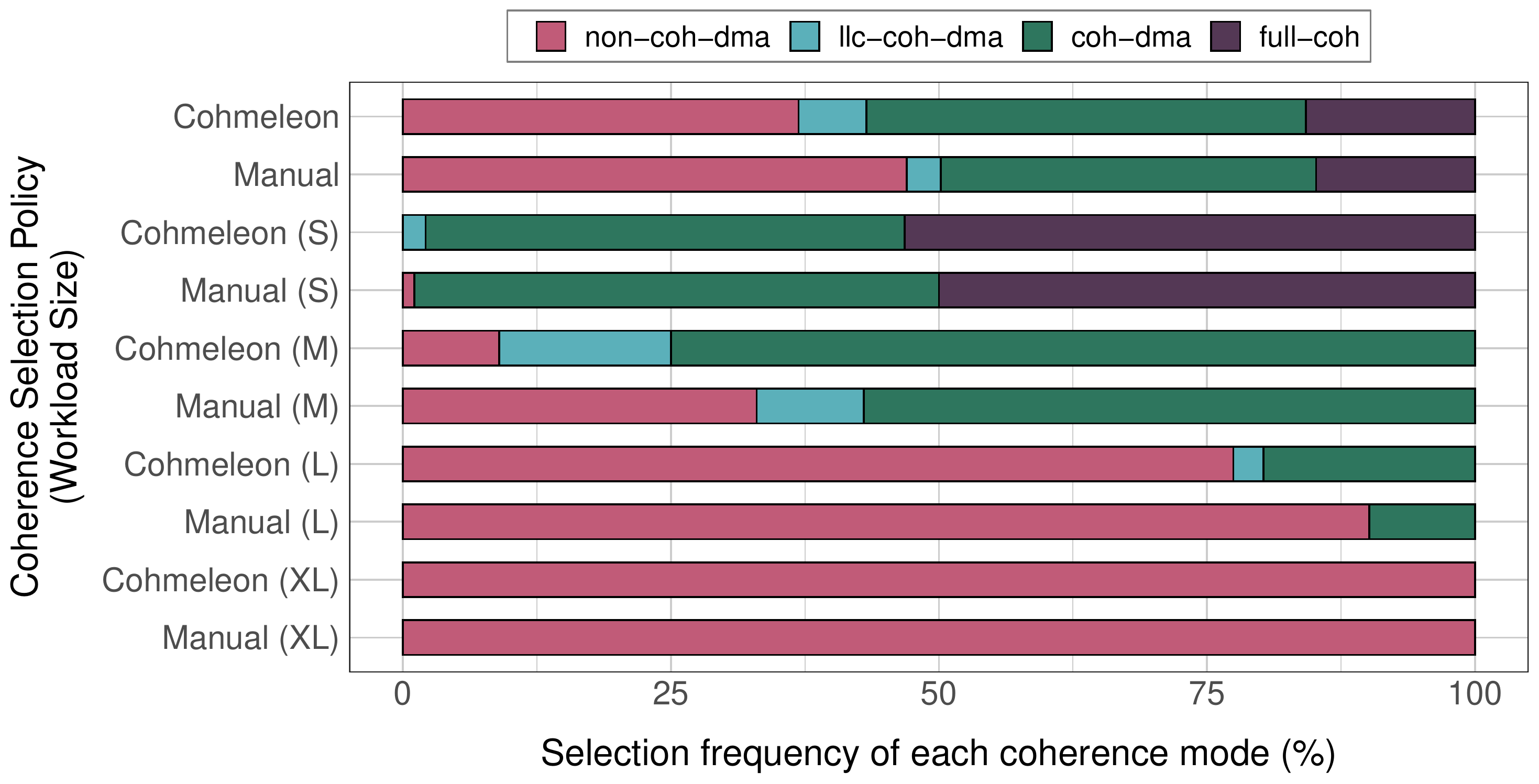}
\centering
\vspace{-0.6cm}
\caption{Breakdown of coherence decisions.}
\label{fig:reward_coh}
\vspace{-0.5cm}
\end{figure}

\textbf{Coherence Breakdown.}
In \figurename~\ref{fig:reward_coh}, we present the breakdown of coherence
decisions by \cohmeleon and the manually-tuned algorithm. We report both the
total breakdown and the breakdown for each category of workload size. Across
all invocations, we see a heavy reliance on the \emph{coherent DMA} (\emph{coh-dma}) and
\emph{non-coherent DMA} (\emph{non-coh-dma}) modes.
Broadly, \cohmeleon seems to learn a policy that results in a breakdown
of coherence modes similar to that of the manually-tuned algorithm.
However, across all categories (except XL workloads), \cohmeleon relies less
upon the \emph{non-coherent DMA} and more upon \emph{coherent DMA} and
\emph{LLC-coherent DMA} (\emph{llc-coh-dma}) than the manual algorithm. The
\emph{non-coherent DMA} mode typically results in the highest number of
off-chip memory accesses for workloads that fit on-chip.  Due to its
bi-objective reward, \cohmeleon tries to avoid this selection in such a situation.

\begin{figure}[t]
\includegraphics[width=\linewidth]{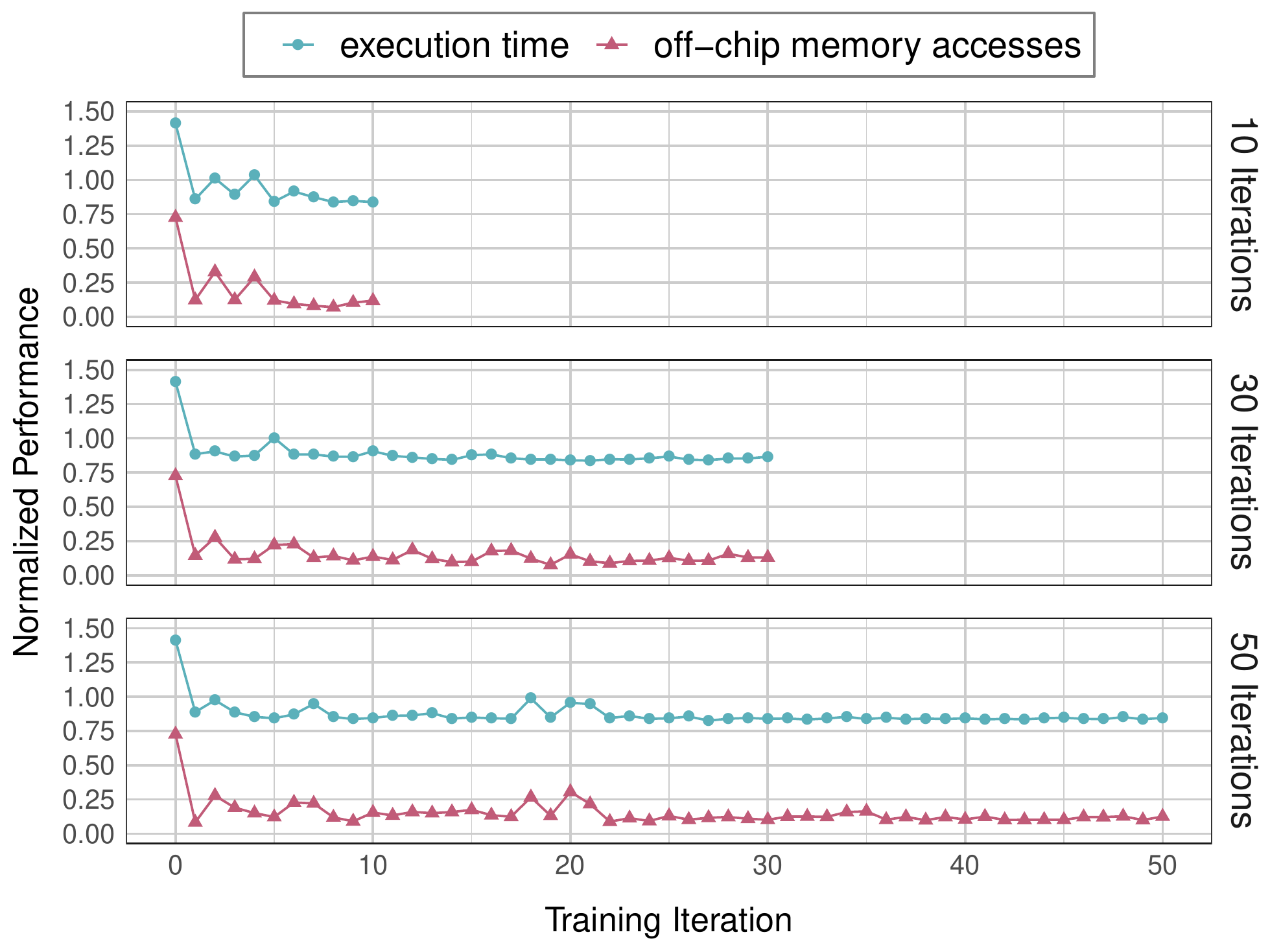}
\centering
\vspace{-0.6cm}
\caption{Performance over training iterations.}
\label{fig:training}
\vspace{-0.8cm}
\end{figure}

\textbf{Training Time.}
 To evaluate the effects of training time on \cohmeleon, we run a series of
 experiments in which we evaluate the performance of the RL 
 model after each training iteration.  We alternate the training of
 \cohmeleon on one iteration of an instance of the evaluation application with
 testing the resulting model on a different instance of the evaluation
 application. \rev{Both instances of the application contain several
 hundred accelerator invocations and are designed to be as diverse as possible in
 terms of operating conditions}. We repeat this experiment for 10, 30, and 50 total
 iterations. Each trial initializes $\epsilon$
 (exploration rate) to $0.5$ and $\alpha$ (learning rate) to $0.25$. Each value
 is decayed linearly to 0 over the course of training, thus making the decay
 rates different for each number of iterations.  \figurename~\ref{fig:training}
 shows the performance after each iteration, reported as the geometric mean
 over all phases of the performance normalized to the \emph{Fixed non-coherent DMA}
 policy. Iteration 0 is the performance of an untrained model, equivalent to the \emph{Random} policy. 
 We see a quick
 drop-off in execution time and off-chip memory accesses after just one
 training iteration for all models. 
 \cam{A training iteration includes over 300 accelerator invocations, which
provide for enough exploration to immediately learn an improved policy.  We see
some oscillation in the results for the next few iterations, as the models
continue to explore the state space with different actions. All models reach
approximately the same performance at the end of training.  We conclude that ten
iterations are sufficient to achieve near-optimal results.}

\begin{figure*}[t]
\includegraphics[width=0.95\linewidth]{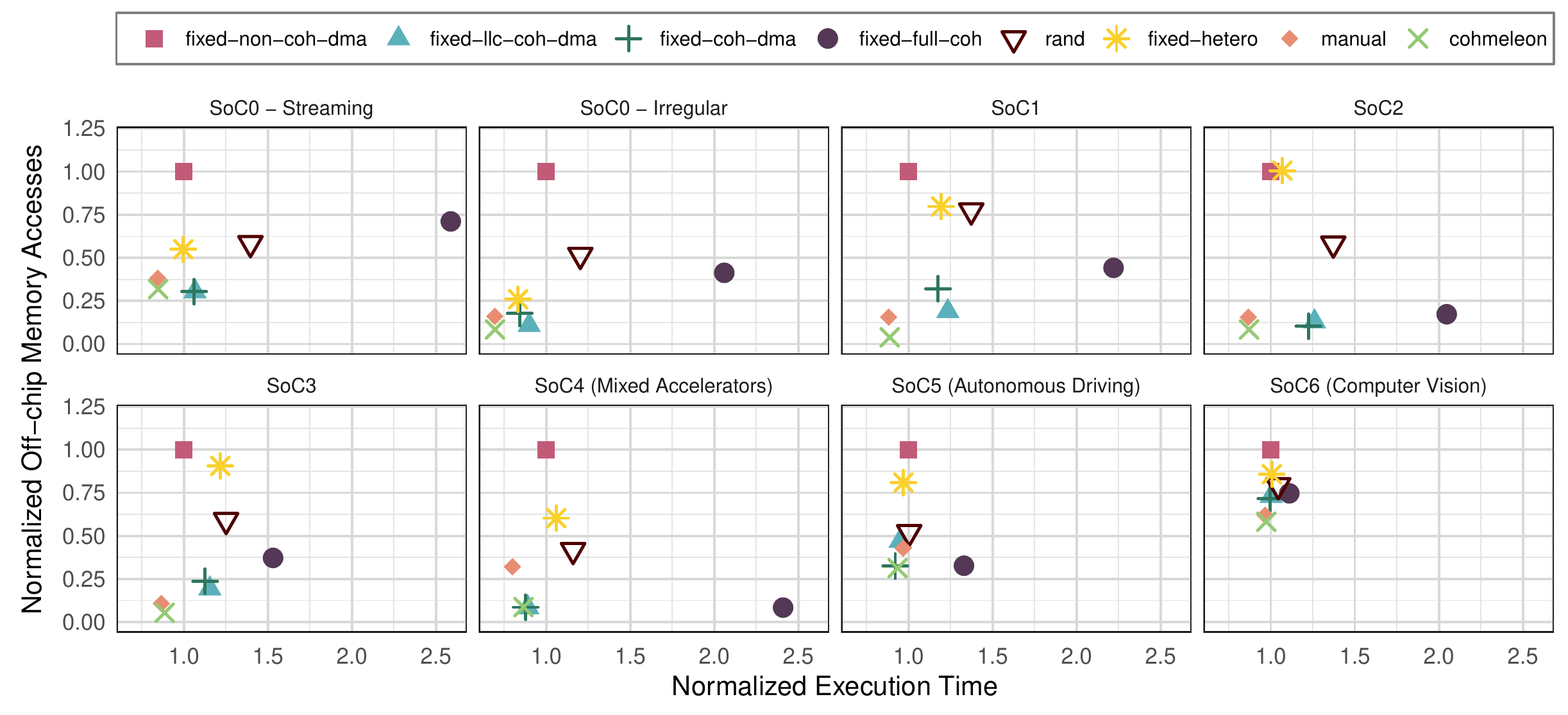}
\centering
\vspace{-0.3cm}
\caption{Performance across SoCs with different accelerators and architectural parameters.}
\label{fig:scatter}
\vspace{-0.35cm}
\end{figure*}

\textbf{Additional SoCs.} We repeat our experiments on eight different SoC
configurations, utilizing the SoCs from Table \ref{tab:synthsocs} to verify
that \cohmeleon is effective with different architectural parameters. On each
configuration, we use a reward function of 67.5\% execution time, 7.5\%
communication ratio, and 25\% off-chip memory accesses \cam{and train for 10
iterations of the corresponding application.}
These results are reported in \figurename~\ref{fig:scatter}.

First, we reuse the SoC0 layout, but this time we use one set of accelerators
with streaming access patterns and another set of accelerators that have
irregular access patterns. We observe that the \textit{Fixed non-coherent DMA} has
the best execution time among the fixed homogeneous policies for streaming
accelerators, whereas for irregular accelerators the \textit{Fixed LLC coherent
  DMA} and \textit{Fixed coherent DMA} policies achieve better
execution time with fewer off-chip memory accesses. This clearly shows once
again that the communication patterns can affect the optimal coherence choice. 
In both of these configurations, \cohmeleon and the manual algorithm 
achieve the best execution time, while \cohmeleon slightly lowers off-chip
memory accesses.

We also utilize SoC1, SoC2, and SoC3, each configured with a set of
accelerators with mixed properties. We see the same ordering among the fixed
homogeneous policies, but with some substantial differences in their relative
performance.  (i.e. \emph{Fixed fully-coherent} has roughly $2.2\times$ the
execution time of \emph{Fixed non-coherent DMA} on SoC1, but only $1.5\times$ on
SoC3). The performance of \cohmeleon and the manual
algorithm remains the best across these SoCs.

Finally, we perform the same experiments on the Case Study SoCs, i.e. SoC4,
SoC5, and SoC6. On SoC5 and SoC6, we observe that there is much less variability
among the performances of the fixed modes. Because the accompanying applications
only invoke accelerators in ways that are appropriate for the corresponding
real-world application, there is less diversity in terms of the characteristics
of the applications. \cohmeleon still achieves optimal
or near-optimal performance across these SoCs. In contrast, the manual algorithm
has suboptimal performance on SoC5, showing it is not capable of generalizing
optimally to all SoCs. 
 
Given these observations, it is clear that these SoC configurations have
different communication properties due to their set of accelerators and
architectural parameters. Nonetheless, we see that \cohmeleon can achieve
the minimal or near-minimal value for both execution time and off-chip memory
accesses across all experiments.  \cohmeleon improves as there is more diversity in
performance across the coherence modes. This is intuitive, as there are
opportunities to improve performance by exploiting an intelligent decision.
Across all SoC configurations, \cohmeleon gives an average speedup of 38\% with
a 66\% reduction in off-chip memory accesses when compared to the five fixed
policies. 

\textbf{Cohmeleon Overhead.}  We measured the
fraction of the total execution time caused by \cohmeleon's status
tracking, computation, and decision-making, which is included in all prior results. 
For small (16KB) workloads, the
overhead is between 3 and 6\% of the total execution time. This value drops as
the workload size increases and the accelerators have longer execution
times. For large (4MB) workloads, the overhead is safely below $0.1\%$,
a negligible value.
\vspace{-0.3cm}

\section{Related work}
\label{sec:related}

\textbf{Comparing Accelerator Coherence Modes.}
There are only a few studies that compare the cache-coherence options for
accelerators, as we did in Section~\ref{sec:motivation}.
Kumar \textit{et al.} propose a fully-coherent approach based on a
timestamp-based coherence protocol~\cite{timestampcoh_tpds92} and compare it
with classic fully-coherent and coherent-DMA solutions~\cite{fusion_isca15}.
Shao \textit{et al.} investigate the non-coherent and fully-coherent
modes~\cite{shao_micro16}.
Cota \textit{et al.} evaluate LLC-coherent and non-coherent accelerators~\cite{cota_dac15}.
These works focus on the simulation of simple SoCs with a few accelerators.
We prototype NoC-based SoCs with up to 16 accelerators using an FPGA-based setup
to run complex real-size applications that manage multiple accelerators on top
of Linux.
The setup of Giri \textit{et al.} is similar to ours, but they did not evaluate
all four cache-coherence modes~\cite{giri_ieeemicro18, giri_nocs18}.

\textbf{Heterogeneous Accelerator Coherence Modes.}
Bhardwaj \textit{et al.}  propose a machine learning approach to assign an
optimal coherence mode to each accelerator at design time~\cite{bhardwaj_cal19,bhardwaj_islped20}.
Giri \textit{et al.} propose a manually-designed algorithm for deciding the
cache-coherence mode at runtime, based on the system
status~\cite{giri_aspdac19}.  \rev{These approaches do not handle all four
  cache-coherence modes, are not updated online, \cam{and require specific
    tuning for the target architecture}; further, the latter is not a
  learning-based approach.}

\textbf{Multi-chip Accelerator Coherence.}
Cache coherence is relevant also for systems where the accelerators live on
their own chip and communicate with a host processor core via an I/O interface,
such as PCIe.
Industry examples of cache-coherent chip-to-chip interconnect for accelerators
include CCIX~\cite{ccix_intro, ccix17}, CXL~\cite{cxl}, IBM CAPI~\cite{capi_ibmjrd15},
OpenCAPI~\cite{opencapi16}, Arteris NCore~\cite{arteris_white_paper_2016} and ARM
CoreLink~\cite{arm_corelink}.
Similarly to the single-chip case, they offer multiple options for handling the
accelerators' cache coherence.
Hence, our approach could be applied also to multi-chip systems.

\textbf{Cache Bypassing.}
\cam{The coherence modes classified in Section~\ref{sec:background} differ based
on the degree of hardware coherence and cache bypassing. The
cache bypassing for fixed-function LCAs has task granularity and doesn't
require modifications to the cache hierarchy. Differently, because of the
programmable nature of GPUs, the literature proposes a
variety of GPU-specific cache bypassing techniques with instruction granularity
and that require modification either to the compiler or the cache
hierarchy~\cite{li_ics15, tian_gpgpu15, xie_hpca15, li_sc15, xie_iccad13}.}

\textbf{RL for SoC Control Problems.}
Although \cohmeleon is the first work using online learning to
orchestrate accelerator coherence, many online learning methods have been
proposed in various application domains.
Liu \textit{et al.} propose a dynamic
power manager based on Q-learning that does not require any prior knowledge of
the workload~\cite{Liu2010Enhanced}. Gupta \textit{ et al.} present a deep
Q-learning approach to dynamically manage the type, number, and frequency of
active cores in SoCs~\cite{Gupta2019Deep}. Zheng \textit{et al.} propose an
energy-efficient NoC design with DVFS (dynamic voltage and frequency scaling)
and a per-router Q-learning agent for
selecting voltage/frequency values~\cite{Zheng2019Energy}.
Besides Q-learning and RL, other machine-learning approaches have also been
proposed for system optimization~\cite{donyanavard2019sosa, ding2019generative}.

\vspace{-0.3cm}
\section{Conclusions}
\label{sec:conclusions}
We showed that the performance of \rev{fixed-function LCAs} in SoC architectures
benefits from runtime reconfiguration of their cache-coherence mode.  \cohmeleon
applies reinforcement learning to automatically and adaptively select the
optimal cache-coherence mode at the time of each accelerator's invocation. It
operates in a way that is transparent to the programmer, with negligible
overhead, and without any knowledge about the target accelerators and
architecture.
\cam{We released \cohmeleon and its FPGA-based experimental infrastructure as
  part of the open-source ESP project~\cite{esp}}.
\vspace{-0.5cm}

\begin{acks}
This work was supported in part by DARPA (C\#:FA8650-18-2-7862),
the ARO (G\#:W911NF-19-1-0476), the NSF (A\#:1764000) and the NSF Graduate
Research Fellowship Program.
The views and conclusions expressed are those of the authors and should
not be interpreted as representing the official views or policies of the
Army Research Office, the Department of Defense, or the U.S. Government.
\end{acks}
\newpage
\newpage
\bibliographystyle{ACM-Reference-Format}
\bibliography{paper}

\end{document}